\documentstyle[12pt,aasms4]{article}

\begin{document}

\title{Fluxtube model atmospheres and Stokes V zero-crossing wavelengths}
\author{L.R. Bellot Rubio, B. Ruiz Cobo, M. Collados}
\affil{Instituto de Astrof\'{\i}sica de Canarias, E-38200, La
Laguna (Tenerife), Spain}
\authoremail{lbellot@iac.es, brc@iac.es, mcv@iac.es}

\begin{abstract} 
First results of the inversion of Stokes I and V profiles from
plage regions near disk center are presented. Both low and high
spatial resolution spectra of \ion{Fe}{1} 6301.5 and \ion{Fe}{1} 
6302.5 \AA\ obtained with the Advanced Stokes Polarimeter (ASP) 
have been considered for analysis. The thin flux tube approximation,  
implemented in an LTE inversion code based on response functions, 
is used to describe unresolved magnetic elements. The code allows 
the simultaneous and consistent inference of all atmospheric 
quantities determining the radiative transfer with the sole
assumption of hydrostatic equilibrium. By considering velocity 
gradients within the tubes we are able to match the full 
ASP Stokes profiles. The magnetic atmospheres derived from the 
inversion are characterized by the absence of significant motions 
in high layers and strong velocity gradients in deeper layers. 
These are essential to reproduce the asymmetries of the observed 
profiles. Our scenario predicts a shift of the Stokes V 
zero-crossing wavelengths which is indeed present in observations 
made with the Fourier Transform Spectrometer.  
\end{abstract}


\section{Introduction}
A remarkable, still unexplained feature of the spectra emerging 
from facular and network regions in the solar photosphere is the 
conspicuous asymmetry exhibited by Stokes V profiles. At disk 
center, the peak and absolute area of the blue lobe are larger 
than those of the red lobe for the majority of \ion{Fe}{1} lines 
(Solanki \& Stenflo 1984). In addition, the red lobe has a more 
extended wing than its blue counterpart. A detailed analysis of 
the properties of high spatial resolution Stokes V spectra in 
plage regions can be found in Mart\'{\i}nez Pillet et al.\ (1997). 

The search of the processes that give rise to the area asymmetry 
of Stokes V has been one of the most vivid discussions of recent 
times in solar physics. Illing et al.\ (1975) first suggested that 
gradients of velocity and magnetic field along the line of sight 
may produce asymmetric V profiles. Originally, this mechanism was
put forward to explain broad-band measurements of circular polarization 
in sunspots. Later, a series of papers refined this idea and settled 
down the physics involved (van Ballegooijen 1985; S\'anchez Almeida 
et al.\ 1988; Grossmann-Doerth et al.\ 1988). In the current picture, 
the area asymmetry is produced by the combined (but otherwise {\em 
spatially separated}) gradients of magnetic field and velocity that 
photons traversing the tubes encounter at the boundary layer. These 
gradients are generated by the expanding walls of magnetic elements 
embedded in field-free surroundings. The canopy mechanism, however, 
does fail to explain the peak asymmetry of Stokes V spectra. 
Observationally, the peak asymmetry turns out to be several times 
larger than the area asymmetry. All attempts to reproduce the 
observations have invariably led to ratios of peak to area 
asymmetries of the order of unity. 
 
A number of other scenarios have been invoked to solve this problem. 
They include NLTE effects (Kemp et al.\ 1984; Landi Degl'Innocenti 
1985), linear oscillations (Solanki 1989) or longitudinal waves
(Solanki \& Roberts 1992) within the magnetic elements, non-linear 
oscillations (Grossmann-Doerth et al.\ 1991) and, more recently, 
micro-structured magnetic atmospheres (S\'anchez Almeida et al.\ 1996). 
None of these mechanisms has proven yet to be able to reproduce the 
observations. Non-linear, high-frequency oscillations inside the tubes 
have never been detected in spite of the observational efforts. Time 
series of Stokes V profiles at disk center analyzed by Fleck (1991), 
for example, did not reveal such motions. Micro-structured magnetic 
atmospheres, on the other hand, seem promising but still need further 
development. 

The present state-of-the-art can be summarized as follows: no 
physical process is known to be able to generate a peak asymmetry 
significantly larger than the area asymmetry exhibited by Stokes V 
spectra in plage and network regions at disk center. In fact, 
some of the mechanisms above were proposed, without much success, 
as a very last attempt to explain the observations. It is our 
belief, however, that velocity gradients have not been completely 
exploited yet. A natural extension of the canopy mechanism might
include mass motions in the magnetized plasma. Empirical evidence 
of such motions has been found repeatedly (Solanki 1986, 1989; 
S\'anchez Almeida et al.\ 1990), but stationary flows were ruled 
out already in the earliest analyses as the dominant source of the 
asymmetries because they would generate Doppler shifts of the 
Stokes V zero-crossing wavelengths larger than the reported upper 
limit of $\pm 250$ m s$^{-1}$ (Solanki 1986). Some preliminary 
calculations, often involving one dimensional models, have confirmed 
this result (e.g., Solanki \& Pahlke 1988). However, it could be 
possible that the small shifts of the zero-crossings are effectively 
produced by velocity gradients in the magnetic atmosphere. These 
gradients might explain the marginal dependence between the observed 
shifts and line strength that various analyses have revealed 
(Solanki 1986; S\'anchez Almeida et al.\ 1989). If this is the 
case, stationary flows should be reconsidered again as a possible 
source of the asymmetries.

In order to proceed further, reliable 2-D fluxtube atmospheres 
have to be used. Unfortunately, no such models exist. Numerical 
simulations (e.g.\ Steiner et al.\ 1995) are more intended to 
understand the interaction between magnetic elements and their 
convective surroundings than to explain the observed spectra. 
For this reason it is not strange that, despite their important 
results, they are not able to reproduce the asymmetries. On the 
other hand, empirical models fail to match the observations 
because they neglect the role of velocity gradients to 
diminish the complexity of the problem. Indeed, extracting 
the information contained in the Stokes spectrum is hampered by 
the intricate non-linear dependences of the radiative 
transfer equation on the various quantities defining both the 
thermodynamical and magnetic properties of the atmospheres. 
The trial-and-error method does not work in this case, and so 
inversion techniques come into play. Here we follow an empirical
approach to investigate the capabilities of velocity-based 
mechanisms. As a first step, a new LTE inversion code of the 
radiative transfer equation particularized to the case of thin 
flux tubes has been developed. Applied to real observations, 
it carries out a simultaneous inference of the whole set of model 
parameters which reproduce the observed Stokes spectrum, thus 
ensuring self-consistency. The details of the procedure and 
numerical tests will be presented in a forthcoming paper, although 
a brief description of the method is given below. 

This letter reports on fluxtube model atmospheres derived from 
the inversion of high spatial ($\sim 1"$) and temporal ($\sim 4$ 
sec) resolution Stokes I and V spectra of \ion{Fe}{1} 6301.5 and 
6302.5 \AA\ from plage regions at disk center obtained with the 
Advanced Stokes Polarimeter (ASP, Mart\'{\i}nez Pillet et al.\ 1997). 
For the first time, plausible model atmospheres are found that 
reproduce the whole shape of the ASP \ion{Fe}{1} 6301.5 and \ion{Fe}{1} 
6302.5 \AA\ Stokes spectra to a degree of accuracy never reached before. 
The recovered model atmospheres, in which stationary flows within 
the tubes were allowed, have been used to synthesize the Stokes 
profiles of a large number of spectral lines. The comparison of
their zero-crossing wavelengths to observations made with the 
Fourier Transform Spectrometer (FTS, Stenflo et al.\ 1984) reveals 
that the observed Doppler shifts are not randomly distributed 
and suggests that non-negligible mass motions within the 
magnetic elements do exist.

\section{Inversion of Stokes profiles from solar magnetic elements 
and observations}
The code we have developed is a generalization of a previous 
code by Ruiz Cobo \& del Toro Iniesta (1992) which carries out
the inversion of the radiative transfer equation by employing a 
Marquardt non-linear least squares fit. Information about
how the emergent Stokes vector varies when the atmospheric
parameters are changed is provided by the so-called response 
functions (Ruiz Cobo \& del Toro Iniesta 1994). The use of
response functions (RFs) accelerates considerably the iterative 
scheme, making it fully automatic. 

The extension of the code to the case of unresolved magnetic
elements implies the adoption of a model and the calculation 
of its RFs. We have adopted the thin flux tube approximation 
to describe small scale magnetic structures.  Axisymmetric 
thin tubes are characterized by the radial constancy of the 
physical parameters within the tube and satisfy hydrostatic 
equilibrium, horizontal pressure balance and magnetic flux 
conservation. The model consists of two homogeneous atmospheres, 
namely the tube itself and the non-magnetic surroundings. To 
take proper account of the geometry, the emergent Stokes 
spectrum is computed as an average of vertical rays that pierce 
the tube at different radial distances. The inversion problem 
is kept tractable by considering the non-magnetic atmosphere 
horizontally constant, which is a reasonable assumption for 
high spatial resolution spectra. The free parameters of the 
model are the temperature and line of sight velocity stratifications 
inside and outside the tube, constant microturbulent velocities in 
both atmospheres, the radius and the magnetic field strength of the 
tube at a given height, the external gas pressure at that level 
and the same height-independent macroturbulence for both atmospheres. 
Contributions due to stray light have been included as well. 
The RFs of the Stokes spectrum emerging from such a model were 
derived and their basic properties discussed by Bellot Rubio, 
Ruiz Cobo \& Collados (1996).

The inversion code has been applied to averaged and spatially 
resolved ASP Stokes spectra of \ion{Fe}{1} 6301.5 and \ion{Fe}{1} 
6302.5 \AA\ from plage regions in NOAA 7197 (Mart\'{\i}nez Pillet 
et al.\ 1997). Typical signal-to-noise ratios are of the order 
of 500-1000 in the continuum of Stokes I. The average profiles 
have been constructed from individual spectra whose degree of 
polarization integrated over a band in 6302.5 \AA\ is larger 
than 0.4$\%$. As a consequence, they are more representative 
of strong magnetic fields and/or high filling factor regions. 

For comparison purposes, very high signal-to-noise, low spatial and 
temporal resolution FTS spectra from plage regions at disk center will 
be employed (Stenflo et al.\ 1984). These observations provide a huge 
number of lines for analysis which will allow us to draw statistically 
significant conclusions. 

\section{Results}
The inversion scheme iteratively modifies a guess atmosphere
until the best fit between synthetic and observed Stokes spectra
is reached. The instrumental profile of the spectrograph 
(Mar\-t\'{\i}\-nez Pillet et al.\ 1997) is taken into account 
to eliminate spurious asymmetries and Doppler shifts. For 
both the magnetic interior and the surroundings, the HSRA 
model atmosphere (Gingerich et al.\ 1971) has been adopted 
initially. The magnetic field strength and the radius of the tube 
at the base of the photosphere (placed arbitrarily at $z = -122$ km) 
were taken to be 1500 G and 50 km. For both atmospheres the initial 
microturbulence was set at 0.6 km s$^{-1}$, the macroturbulence at 
1 km s$^{-1}$ and the line-of-sight velocity at 1 km s$^{-1}$. 
This particular choice of the guess atmosphere does not 
influence any of the results of this work, as different 
initializations lead to the same final models. 

Figure 1 shows the average ASP Stokes I and V spectra of 
\ion{Fe}{1} 6301.5 and 6302.5 \AA\ inverted as described above,  
and the results of the inversion. Error bars have been estimated 
by assuming the independence of all model parameters. Under 
this hypothesis, confidence limits in the inferred parameters 
turn out to be inversely proportional to linear combinations 
of the RFs. We emphasize, however, that only formal errors 
are taken into account; the possible inadequacies of the thin 
flux tube model, the approximations and the numerical limitations 
of the inversion algorithm are other sources of error whose influence 
is difficult to assess. In any case, the excellent fit obtained 
means that the simple thin tube model is able to mimic real 
observations. At this juncture we want to remark that, even
for only two spectral lines, none of the previous attempts 
to reproduce the shape of Stokes profiles coming from active 
regions outside sunspots has been as successful as the 
inversions reported here. Also, it is important to note that 
we have not antisymmetrized the V profiles.

\placefigure{fig1}

The relative Stokes V area asymmetries $\delta A$ of the average 
ASP \ion{Fe}{1} lines at 6301.5 and 6302.5 \AA\ turn out to be 
$2.8\%$ and $4.0\%$, whereas the relative amplitude asymmetries 
$\delta a$ amount to $10.9\%$ and $9.9\%$ (for a definition of 
$\delta A$ and $\delta a$ see, e.g., Solanki 1989). These values 
are correctly reproduced by the combined action of downflows 
inside and outside the tubes. The synthetic spectra of 6301.5 
and 6302.5 \AA\ presented in Fig.\ 1 have $\delta A = 3.2\%$ 
and $\delta A = 2.2\%$. The corresponding values of $\delta a$ 
are $8.1\%$ and $7.1\%$. In order to explain the full Stokes profiles 
and, in particular, their asymmetries, the velocity gradient 
recovered by the inversion procedure in the magnetic atmosphere 
is essential. This most important conclusion has been drawn after 
a large number of inversions in which no velocities within the 
tubes were allowed. 

The basic features of the model atmospheres that fit the 
average Stokes spectrum are still maintained by those inferred
from the inversion of individual spectra. Figure 2 shows the
Stokes profiles emerging from a number of adjacent pixels in
a plage region of NOAA 7197 and the results of the inversions. 
Non-negligible differences between spectra coming from adjacent 
points exist, specially in Stokes V and the continuum of Stokes 
I. These, however, are reproduced by models that share the same 
basic properties as the atmospheres determined from the average 
spectra. This strengthens the reliability of the procedure 
and validates the models to a large extent.

\placefigure{fig2}
 
We want to remark that the internal downflows derived from
the inversion of observed spectra are compatible, within error 
bars, with null velocities in high atmospheric layers; only in 
deep layers is an abrupt velocity gradient established. This 
implies a different view about how the asymmetries are generated. 
For convenience, we shall make a distinction between the cylinder 
inside the tube and the rest of the magnetic element (the canopy). 
While the canopy gives rise to pronounced positive amplitude and 
area asymmetries, the cylinder produces negative area and positive 
peak asymmetries. This results in a peak asymmetry significantly 
larger than the area asymmetry, in accordance with the observed 
behavior of Stokes V at disk center. The role of the canopy is 
maintained in our scenario, but now the cylinder creates the 
necessary {\em negative} area asymmetry as to compensate the 
excess of area asymmetry produced by the canopy. Gradients of 
velocity ($v$) and field strength ($B$) are cospatial and 
negative within the tube. This leads to the condition $\frac{{\rm d} 
|B|}{{\rm d} z} \frac{{\rm d}v}{{\rm d}z} > 0$ almost everywhere, 
which fully explains the sign of $\delta A$ in the magnetic 
cylinder (Solanki \& Pahlke 1988).

\section{Zero-crossing wavelengths of observed Stokes V profiles} 
It has been argued from long ago that stationary velocity
fields within the magnetic elements cannot be considered
as the main source of the asymmetries of Stokes V because the 
required motions would produce noticeable shifts of their 
zero-crossing wavelengths. An observed upper limit of about 
$\pm 0.25$ km s$^{-1}$, which is only slightly larger than 
the estimated accuracy of current wavelength calibrations, 
has been set in different analyses (Stenflo \& Harvey 1985; 
Solanki 1986). 

The atmospheres presented in Sect.\ 3 successfully reproduce 
the observed profiles of two spectral lines. This means
that both area and amplitude asymmetries, and zero-crossing 
wavelengths are also matched. But, do the recovered velocity 
fields generate Stokes V zero-crossing shifts compatible with 
the observations for a large number of spectral lines? In 
order to check this point we have used the models resulting 
from the inversion of the average profiles and a macroturbulence 
of 1 km s$^{-1}$ to synthesize the Stokes spectra of 92 unblended 
\ion{Fe}{1} lines with different excitation potentials and 
strengths in the range 4600-5450 \AA\/. Their Stokes V 
zero-crossings have then been compared to FTS observations. 
Figure 3 summarizes the analysis and reveals a distinct  
similarity between observed and computed shifts. FTS 
wavelengths are not absolutely calibrated, but in any case 
the correlation found suggests that the velocity gradients 
derived by the inversion procedure are those required by 
the observations to be reproduced. Since the selected lines 
sample the whole photosphere, we conclude that mass motions 
within fluxtubes increase with depth, thus producing larger 
zero-crossing shifts for lines formed at deeper layers. 

\placefigure{fig3}

In view of the evidence presented here we believe that strong 
indications of velocity gradients in the magnetized plasma exist, 
and that these may be responsible for the observed asymmetries 
of Stokes V in plage regions at disk center. Various mechanisms 
might explain the downflows, including convective collapse, gas 
entry into the magnetic structures and viscous drag induced by 
external convective motions (Steiner et al.\ 1995). Certainly, 
the origin of velocity gradients in the magnetic atmosphere 
deserves further theoretical work.

\acknowledgments
We gratefully acknowledge V.\ Mart\'{\i}nez Pillet, B.\ Lites and 
A.\ Skumanich for providing their reduced ASP observations and for 
many useful discussions regarding the data. Thanks are also due to 
S.\ K.\ Solanki, who kindly allowed us to compare the FTS zero-crossing 
shifts with the predictions of our model and made valuable suggestions. 
This work has been funded by the Spanish DGICYT under project PB91-0530.


\newpage
\figcaption[figure1n.eps]{Inversion of average ASP Fe I 6301.5 and 6302.5 
\AA\ spectra. Top panels: observed (open circles) and synthetic (solid line) 
Stokes profiles, and residuals. Bottom panels: models derived from the 
inversion. The magnetic interior and the external surroundings are 
represented by solid and dashed lines, respectively. The best estimates 
for the single-valued parameters of the model turn out to be  $2.0 
\pm 0.1$ km s$^{-1}$ for the macroturbulence; $90 \pm 3$ km, $2050 \pm 
50$ G and $(3.47 \pm 0.06) \times 10^5$ dyn cm$^{-2}$ for the radius of 
the tube, the magnetic field and the external gas pressure at the base 
of the photosphere; $61 \pm 2 \%$ for the stray light contamination and 
null microturbulences. \label{fig1}}

\newpage
\figcaption[figure2n.eps]{Inversion of 16 high spatial resolution ASP Fe 
I 6301.5 and 6302.5 \AA\ spectra with maximum V signals larger than 0.05 
and maximum Q and U signals lower than 0.002 (in units of the continuum 
of Stokes I). Top panels: observed spectra. The residuals below are 
representative of a typical fit. Bottom panels: models derived from the 
inversion. Solid and dashed lines represent the magnetic interiors and 
the external surroundings, respectively. The inferred macroturbulence 
ranges from 1.6 to 1.9 km s$^{-1}$, the internal microturbulent velocity 
from 0 to 0.9 km s$^{-1}$, and the stray light contamination from 21 to 
30 $\%$. The external microturbulence is always null. At the base of 
the photosphere, the radii of the tubes range from 95 to 107 km, the 
magnetic field strength from 2070 to 2430 G, and the external gas 
pressure from $3.43 \times 10^{5}$ to $3.80 \times 10^{5}$ dyn 
cm$^{-2}$. \label{fig2}}

\newpage
\figcaption[figure3n.eps]{Predicted and observed Stokes V zero-crossing 
shifts for 92 unblended Fe I lines in the range 4600-5450 \AA\/. 
Laboratory wavelengths from Nave et al.\ (1994) have been adopted 
because solar wavelengths are affected by convective blueshift. For 
comparison purposes, a straight line with slope equal to unity has 
been added. Dotted lines indicate the uncertainty of laboratory 
wavelengths and lie $\pm 0.15$ km s$^{-1}$ apart. 
\label{fig3}}

\end{document}